\documentclass[12pt]{article}

\usepackage{theorem,amssymb,amsbsy,latexsym,amsmath,bbm}

\theoremstyle{plain}

\newcommand{\RR}{\Delta} 
\renewcommand{\c}{c}
\newcommand{\cB}{c_B}

\newcommand{\nd}{{\phantom\dag}}

\newcommand{\cH}{{\cal H}}

\newcommand{\sgn}{\ensuremath{\textrm{sgn}}}

\newcommand{\ee}[1]{{\rm e}^{#1}}
\newcommand{\ii}{{\rm i}}
\newcommand{\dd}{{\rm d}}

\newcommand{\Ref}[1]{(\ref{#1})}

\newcommand{\R}{{\mathbb R}}

\newcommand{\eq}{\begin{equation}}
\newcommand{\eqend}{\end{equation}}
\newcommand{\eqa}{\begin{eqnarray}}
\newcommand{\nonueqa}{\begin{eqnarray*}}
\newcommand{\eqaend}{\end{eqnarray}}
\newcommand{\nonueqaend}{\end{eqnarray*}}
\newcommand{\nonu}{\nonumber \\ \nopagebreak}
\newcommand{\bma}[1]{\begin{array}{#1}}
\newcommand{\ema}{\end{array}}
\newcommand{\bc}{\begin{center}}
\newcommand{\ec}{\end{center}}
\newcommand{\sct}[1]{\noindent {\bf {#1}}}

\newcounter{saveeqn}
\newcounter{App} %\setcounter{App}{0}
\newcommand{\app}{%
\stepcounter{App}%
\setcounter{saveeqn}{\value{equation}}%
\setcounter{equation}{0}%
\renewcommand{\theequation}{\Alph{App}\arabic{equation}} }
\newcommand{\appende}{%
\setcounter{equation}{\value{saveeqn}}%
\renewcommand{\theequation}{\arabic{equation}}  }
\newcounter{asaveeqn}

\begin{document}
\begin{flushright}
ESI Preprint 1424\\
math-ph/0401003v2\\
\vspace{.4cm}

May 12, 2004 
\end{flushright}
\vspace{.4cm}

\begin{center}

{\Large\bf Exact solution of a 1D quantum many-body system with
momentum dependent interactions\footnote{Corrected version of an
article with the same title in J.\ Phys.\ A: Math.\ Gen.\ {\bf 37} No
16 (2004), 4579--4592.}}

\vspace{1 cm}

{\large Harald Grosse$^*$, Edwin Langmann$^{**}$, and Cornelius
Pauf\/ler$^{**}$}\\
\vspace{0.3 cm} {\em $^{*}$ Institut f\"ur Theoretische Physik, Universit\"at Wien,
Boltzmanngasse 5, A-1090 Wien, Austria}\\
\vspace{0.3 cm} {\em $^{**}$ Mathematical Physics, Department of
Physics, KTH, AlbaNova, SE-106 91 Stockholm, Sweden}\\
\end{center}

\begin{abstract}
We discuss a 1D many-body model of {\em distinguishable} particles
with local, momentum dependent two-body interactions. We show that the
restriction of this model to fermions corresponds to the
non-relativistic limit of the massive Thirring model. This fermion
model can be solved exactly by a mapping to the 1D boson gas with
inverse coupling constant. We provide evidence that this mapping is
the non-relativistic limit of the duality between the massive Thirring
model and the quantum sine-Gordon model. We also investigate the
question if the generalization of this model to distinguishable
particles is exactly solvable by the coordinate Bethe ansatz and find
that this is not the case.

\end{abstract}

\sct{1. Introduction.} In this paper we present, discuss, and solve a 
non-relativistic many-body system of particles moving in one space
dimension (1D) and interacting with a particular local, momentum
dependent two-body potential. As we will explain, this model is the
natural fermion-analog of the 1D boson gas.

The 1D boson gas is one of the famous exactly solvable many-body
models. It describes non-relativistic bosons moving in 1D and
interacting with delta-function two-body interactions, and it was
solved by Lieb and Liniger a long time ago \cite{LL} (a nice textbook
discussion of this model and its solution can be found in Chapter I of
reference \cite{Korepin}). We mention only in passing the considerable
recent interest by experimental physicists triggered by a proposal of
an experimental realization of this model in \cite{O}.

As mentioned, the particles in the 1D boson gas interact via a
delta-function interaction. Due to the Pauli principle, this kind of
interaction is trivial for fermions, and thus interesting fermion
models with such an interaction require additional internal degrees of
freedom \cite{G,FL}. Our 1D many-body model is without internal
degrees of freedom and with a particular local, translation invariant
interaction which is non-trivial for fermions. It is defined by the
following Hamiltonian ($\partial_{x_j}\equiv \partial/\partial x_j$),
\eq H = -\sum_{j=1}^N \partial_{x_j}^2 + 2\lambda \sum_{j<k}
\Bigl(\partial_{x_j} -\partial_{x_k} \Bigr) \delta(x_j-x_k) \Bigl(
\partial_{x_j} -\partial_{x_k} \Bigr) , \label{H} \eqend
with an arbitrary number $N$ of particles moving on the real line,
$-\infty <x_j <\infty$ (we will also mention some generalizations of
our results to an interval of length $L$ with periodic or
anti-periodic boundary conditions, $0\leq x_j \leq L$); the real
parameter $\lambda$ determines the coupling strength. Note that the
interactions depend not only on the particle distance $x_j-x_k$ but
also the momentum difference $\hat p_j -\hat p_k \equiv -\ii
(\partial_{x_j}- \partial_{x_k})$. As we will see (Paragraph~2), due
to this the Pauli principle is circumvented: this interaction is
non-trivial on fermion wave functions, while it is trivial on boson
wave functions. We will derive this model as non-relativistic limit of
the massive Thirring model \cite{Th} (Paragraph~3), in the same way as
the boson gas can be obtained as non-relativistic limit of
$\phi^4$-theory in 1+1 dimensions (see Appendix~B.2). We find that
this fermion model can be solved exactly by mapping it to the 1D boson
gas with the coupling replaced by its inverse (Paragraph~4; as we will
explain, this result is equivalent to the duality observed previously
in \cite{CS}). This relation between our fermion model and the 1D
boson gas is reminiscent to the famous duality between the massive
Thirring model and the quantum sine-Gordon model \cite{C}, and we will
present arguments that it actually is the non-relativistic limit of
the latter (Paragraph~5). A natural question is if the generalization
of this model to distinguishable particles remains exactly solvable by
the coordinate Bethe Ansatz, but, as we shortly discuss in
Paragraph~6, this is not the case.

Since the massive Thirring model is known to be integrable (in certain
formal meanings of this word), it is perhaps not too surprising that
its non-relativistic limit in equation \Ref{H} is exactly solvable. It
thus is worth recalling that, despite of various interesting partial
results \cite{BT,Smirnov}, the Thirring model has not been solved in
full detail. It thus is interesting that its non-relativistic limit
can be solved and studied by the much simpler methods which have been
developed for the 1D boson gas.

In our derivations of non-relativistic limits in Paragraph~3 we start
with the formal definition of the quantum massive Thirring model,
perform expansions in $1/(\mbox{mass}\times \c)$ with $\c$ the
velocity of light, and we use {\em physical} arguments to justify our
ignoring certain terms. In this way we arrive at a non-relativistic
model which is well-defined, in the same spirit as reference
\cite{FW}. It should be possible to make this procedure mathematically
precise using the method proposed in \cite{GGT}.  As mentioned, the
arguments in Paragraph~5 are somewhat heuristic. The other results are
mathematically precise. We tried to keep the main text short, but for
the convenience of the reader we included two appendices: In
Appendix~A we give a complimentary treatment of the singular
interaction in our model, Appendix~B contains details of our
non-relativistic limits.

\bigskip

\noindent{\bf Remark:} {\it In the first version of this paper (which
unfortunately has already appeared in J.\ Phys.\ A: Math.\ Gen.) we
presented an argument which seems to show that the generalized model
defined by he Hamiltonian in equation \Ref{H} and for distinguishable
particles is exactly solvable by the coordinate Bethe Ansatz \cite{Y}.
This argument was incorrect.

The mistake leading to our wrong conclusions was a subtle point in
Yang's arguments \cite{Y}, and we were mislead by one of our sources
to use an unfortunate notation covering up this point. To be more
specific: while equation (C4) in the published version of this paper
is not incorrect, it is misleading the way it is written, and our
interpretation spelled out in Appendix C.1, Remark 2, led us astray:
the twist we proposed to save the validity of the Yang-Baxter
relations leads to an inconsistency at another point which we missed.
The correct interpretation of this equation is now given in section 6.

It is worth mentioning that we discovered this mistake when we tried
to generalize our result to other types of singular interactions,
checking the validity of the Bethe Ansatz for the three particle case
directly using symbolic computer programs written in MAPLE and
MATHEMATICA (to be sure two of us wrote independent programs). While
these programs nicely confirmed Yang's result for the delta
interaction case, they showed that the coordinate Bethe Ansatz for our
model is consistent if and only if the eigenfunction either has boson
or fermion statistics.}

\bigskip

\sct{2. Two particle case.} To get a physical understanding of our
model it is instructive to first consider the two-particle case
$N=2$. Introducing $x=x_1-x_2$ and ignoring the trivial center-of-mass
motion, $H$ in equation \Ref{H} reduces to the following simple
Hamiltonian,
\eq h = -\partial_x^2 + 4\lambda \partial_x \delta(x)\partial_x
\label{H2} , \eqend
whose eigenfunctions $\chi(x)$, $x\in \R$, are defined by satisfying
$(\partial^2_x + E)\chi(x)=0$ for $x\neq 0$ and the following boundary
conditions,
\eqa \chi'(0^+ )-\chi'(-0^+) &=& 0\nonu \chi(0^+ ) -\chi(-0^+) &=&
4\lambda \chi'(0^+) , \label{bc} \eqaend
with the prime indicating differentiation. Indeed, these are the
boundary conditions obtained by integrating $h\chi=E\chi$ twice: first
from $x=-0^+$ to $x>0$ where $\chi'(0)$ is interpreted as the average
of the left- and right derivative, and then once more from $x=-0^+$ to
$0^+$ (it is instructive to verify this formal argument by checking
that the solutions below indeed satisfy $h\chi = E\chi$). The
solutions of this are obtained by simple computations,
\eqa \chi_+(x) &=& \cos(kx)\nonu \chi_-(x) &=&
\frac{\sin(kx)}{2\lambda k } + \sgn(x) \cos(kx) \label{solution}
\eqaend
with corresponding eigenvalue $E=k^2$. For real $k$ these all are
scattering states, and for $\lambda<0$ there is one additional bound
state for $k=\ii / 2\lambda$ with energy $E=-1/4\lambda^2$. Thus
positive and negative values of $\lambda$ correspond to the repulsive
and attractive cases, respectively. As already mentioned, the boson
wave function $\chi_+$ is unchanged by the interaction, while the
fermion wave function $\chi_-$ is modified, opposite to what happens
for the delta-function interaction. It is worth noting that, in
converting the interaction in equation \Ref{H2} into the boundary
conditions in equation \Ref{bc}, we have used a regularization procedure
which consistently avoids divergences which would occur in a naive
treatment of this singular interaction (this is explained in more
detail in Appendix~A).

In a similar manner one finds that the eigenfunctions $\chi$ of the
Hamiltonian in equation \Ref{H} for arbitrary $N$ are given by the
solutions of $(\sum_j \partial_{x_j}^2+E)\chi(x_1,\ldots,x_N) = 0$ in
all regions of non-coinciding points, together with the following
boundary conditions
\eqa \Bigl( \partial_{x_j}-\partial_{x_k}\Bigr) \chi|_{x_j = x_k+0^+}
&=& \Bigl( \partial_{x_j}-\partial_{x_k}\Bigr)\chi|_{x_j = x_k-0^+}
\nonu \chi|_{x_j = x_k+0^+} - \chi|_{x_j = x_k-0^+} &=& 2\lambda
\Bigl( \partial_{x_j}-\partial_{x_k}\Bigr)\chi|_{x_j = x_k-0^+}
\label{BC} 
\eqaend
(we used that $\partial_{x_j-x_k}=
(\partial_{x_j}-\partial_{x_k})/2$).  It is straightforward to check
that these boundary conditions are trivially fulfilled for all
non-interacting boson eigenfunctions $\chi_+ = \sum_{P\in S_N}
\exp(\sum_j \ii k_{Pj} x_j)$. They are, however, non-trivial for
fermions.

\bigskip

\sct{3. Non-relativistic limit of the massive Thirring model.} We now
derive the non-relativistic limit of the massive Thirring model
\cite{Th} and show that it is identical with the second quantization
of the many-body Hamiltonians in equation \Ref{H}. The Thirring model can
be (formally) defined by the quantum field theory Hamiltonian $\cH =
\cH_0 + \cH_{\rm int}$ where the free part is the usual Dirac
Hamiltonian in 1D,
\eq \cH_0 = \int \dd x\; : (\psi_+^\dag,\psi_-^\dag )\left( \bma{cc}
-\ii \c\partial_x - E_0 & m\c^2 \\ m\c^2 & \ii \c\partial_x - E_0 \ema
\right) \left( \bma{c} \psi^\nd_+ \\ \psi^\nd_- \ema \right):
\label{cH0} \eqend
with $m>0$ the fermion mass, and the interaction is
\eq \cH_{\rm int} = 4g \int\dd x\, : \psi_+^\dag \psi_+^\nd
\psi_-^\dag \psi_-^\nd: \label{cHint} \eqend
(see e.g.\ equation (2.1) in \cite{BT}) with $g$ the coupling constant and
the dots indicating normal ordering; the $\psi^{(\dag)}_\pm\equiv
\psi^{(\dag)}_\pm(x)$ are fermion field operators obeying the usual
canonical anticommutation relations (CAR)
$\{\psi^\nd_\pm(x),\psi^\dag_\pm(y)\} = \delta(x-y)$ etc., and $E_0$
is a parameter allowing us to change the reference energy which we
will fix later to a convenient value. One can diagonalize $\cH_0$ by
Fourier transformation and diagonalization of a $2\times 2$ matrix,
which corresponds to a particular canonical transformation
$(\psi^\dag_+, \psi^\dag_-) \to (\Psi^\dag_+, \Psi^\dag_-)$ (see
Appendix~B.1). We expand in powers of $1/m\c$ and obtain, in position
space,
\eq \psi_\pm = \frac{1}{\sqrt2} \Bigl( \Psi_+ \pm \Psi_- \mp 
\frac{\ii}{2m\c}(\partial_x\Psi_+ \mp \partial_x\Psi_-) +
%\frac{\partial_x^2}{8(m\c)^2}(\Psi_+ \pm  \Psi_-) + 
\ldots \Bigl)
\label{Psipm}
\eqend
and $\cH_0 = \cH_0^+ + \cH_0^-$ with $\cH_0^\pm = \pm \int\dd x\,
:\Psi^\dag_\pm [(m\c^2 \mp E_0) -\partial_x^2 /2m + \ldots
]\Psi^\nd_\pm:$, where the dots are for higher order terms in $1/m\c$.
The positive- and negative states of the non-interacting model are now
decoupled, and it is straightforward to compute the interaction in
terms of the new fields $\Psi_\pm$.  To obtain the non-relativistic
limit we set $E_0=m\c^2$ and assume that $m\c^2$ is large. In this
case we can ignore the negative energy degrees of freedom $\Psi_-$:
the non-interacting ground state is such that all the negative energy
states are filled and the positive energy states empty (Dirac sea),
and due to the large energy gap $2m \c^2$ the interactions involving
the filled states, in particular those across the gap, can be
neglected if one is only interested in the low-energy physics. We thus
drop all terms in the Hamiltonian involving the fields
$\Psi^{(\dag)}_-$, and in leading non-trivial order in $1/m\c$ we
obtain the following Hamiltonian,
\eqa \cH_{\rm non-rel} = \int\dd x\, \frac1{2m} \Psi^\dag
(-\partial_x^2) \Psi + \frac{2g}{(2m\c)^2} : \Bigl(
(\partial_x\Psi^\dag) (\partial_x\Psi) \Psi^\dag \Psi - \Psi^\dag
(\partial_x\Psi) (\partial_x\Psi^\dag) \Psi \Bigr): \label{cHnonrel}
\eqaend
with $\Psi\equiv \Psi_+$ obeying CAR and annihilating the
non-interacting vacuum, $\Psi|0\rangle=0$; we used
$:[\Psi^\dag(x)\Psi(x)]^2:=0$, i.e., the lowest order term vanishes
due to the Pauli principle, and thus the leading non-trivial
interaction involves derivatives.  It is straightforward to verify
that this non-relativistic quantum field Hamiltonian $\cH_{\rm
non-rel}$ is the second quantization of our many-body Hamiltonian $H$
in equation \Ref{H}: for $2m=1$ and $g/(2m\c)^2=-\lambda$, the eigenvalue
equation $\cH_{\rm non-rel}|N\rangle=E|N\rangle$ for $N$-particle
states
\eq |N\rangle = \int\dd^N x \, \chi(x_1,\ldots,x_N)
\Psi^\dag(x_1)\cdots \Psi^\dag(x_N) |0\rangle \eqend
is equivalent to $H\chi=E\chi$. Note that $\lambda<0$ corresponds to
$g>0$, in agreement with what one should have expected from the fact
that the massive Thirring model has bound states for $g>0$ (see Eq.\
(2.15b) {\em ff} in \cite{BT}), whereas the sign of $\lambda$ is such
that the attractive case corresponds to $\lambda<0$ (see Paragraph~2
above).

\bigskip

\sct{4. Solution I: Fermion model.} We now determine all fermion
eigenfunctions $\chi$ of the Hamiltonian in equation \Ref{H}. Due to the
fermion statistics we only need to determine
$\chi=\chi(x_1,x_2,\ldots,x_N)$ in the fundamental wedge
\eq \RR_I:\quad  x_1 < x_2 < \ldots < x_N  . \label{R1} \eqend
For the same reason, the boundary conditions in the first line of Eq.\
\Ref{BC} are automatically fulfilled, and the ones in the second line
simplify to $2\chi|_{x_j=x_k+0^+}= 2\lambda ( \partial_{x_j} -
\partial_{x_k}) \chi|_{x_j = x_k+0^+}$ where we only need to consider
the cases $j=k+1$. Thus the equations determining our eigenfunctions
are $(\sum_j\partial_{x_j}^2 + E) \chi=0$ and
\eqa \Bigl( \partial_{x_{j+1}} - \partial_{x_j} - \frac1{\lambda}
\Bigr) \chi|_{x_{j+1} = x_j+0^+} = 0 .  \eqaend
Comparing with equations (2.1a), (2.4a) in \cite{LL} we see that {\em
these conditions are identical with the ones determining the
eigenfunctions of the 1D boson gas defined by the Hamiltonian
\eq H_B = -\sum_{j=1}^N \partial_{x_j}^2 + 2 \cB \sum_{j<k}
\delta(x_j-x_k) \label{Hp} \eqend
at coupling 
\eq \cB = \frac1{\lambda} \label{dual} \eqend
in the fundamental wedge $\RR_I$}. Since the latter eigenfunctions are
well-known, we can immediately write down all eigenfunctions of our
model
\eq \chi(x_1,x_2,\ldots, x_N) = \prod_{1\leq k<j \leq N} \Big(
\lambda[\partial_{x_j} - \partial_{x_k}] + 1 \Bigr) \det_{1\leq
j,k\leq N}[ \exp( \ii k_j x_k ) ] \eqend
in $\RR_I$, and the corresponding eigenvalues are $E=\sum_j k_j^2$ (this
explicit formula is apparently due to Gaudin \cite{Gaudin}; see
Chapter I in \cite{Korepin}).

In this paper we restrict ourselves to particles moving on the full
line, but it is interesting to note that many of our results can be
extended to the finite interval of length $L$, $0\leq x_j\leq L$, with
periodic or anti-periodic boundary conditions,
\eq \chi(x_1,\ldots,x_N) = \ee{\ii \eta} \chi(x_1+L,\ldots,x_N) , \eqend
and similarly for all other arguments $x_j$, with $\eta=0$ or
$\pi$. Similarly as for the 1D boson gas this yields the following
conditions for the allowed momentum values,
\eq \ee{\ii k_j L} = (-1)^N \ee{\ii \eta}
\prod_{\ell=1}^N\frac{k_j-k_\ell +\ii/\lambda}{k_j-k_\ell
-\ii/\lambda} \eqend
(these are the so-called Bethe equations; see e.g.\ Chapter I in
\cite{Korepin}). Comparing with the Bethe equations for the 1D boson
gas (equation (2.2) in \cite{Korepin}) we see that the duality above
remains true for finite interval if we choose in our model periodic
boundary condition ($\eta=0$) if $N$ is even and anti-periodic
boundary conditions ($\eta=\pi$) if $N$ is odd. In the thermodynamic
limit $L,N\to\infty$ such that $\rho=N/L$ remains finite the
difference in boundary conditions becomes irrelevant, and thus {\em
all thermodynamic properties of our model are the same as the known
thermodynamic properties of the 1D boson gas \cite{YY} at inverse
coupling, $\cB=1/\lambda$}.  It would be interesting to know if there
are any observables which can distinguish these two models.

\bigskip

\sct{5. Non-relativistic limit of the quantum sine-Gordon model.} We
now present evidence that the relation of our fermion model to the 1D
boson gas found above is the non-relativistic limit of the duality
between the massive Thirring model and the quantum sine-Gordon (qSG)
model \cite{C}. In the main text we will argue that the qSG model
reduces to $\phi^4_{1+1}$-theory for large (effective) mass. The
result then follows since $\phi^4_{1+1}$-theory in the
non-relativistic limit is identical with the second quantization of
the 1D boson gas (the details of this latter part of the argument are
deferred to Appendix~B.2).

The qSG model can be formally defined by the Hamiltonian
$\cH_{SG}=\cH^B_0 + \cH^B_{1}$ with the usual free boson Hamiltonian
\eq \cH^B_0 = \frac12  \int\dd x\,:\left( \c^2 \Pi^2 + \phi[-
\partial_x^2 + (m\c)^2 ] \phi \right) : \label{cH0p} \eqend
and the interaction 
\eq \cH^B_{1} = \int\dd x\, : \frac{\alpha}{\beta^2}\left[
1-\cos\beta\phi \right] - \frac{(m\c)^2}2 \phi^2 : \eqend
with boson fields $\phi\equiv \phi(x)=\phi^\dag$ and their conjugate
variables $\Pi= \partial_t\phi/\c^2$ obeying the usual canonical
commutation relations (CCR), $[\Pi(x),\phi(y)] = -\ii\delta(x-y)$
etc.; $\alpha$ and $\beta$ are coupling parameters. It is important to
note that, while the bosons in the qSG model are massless, the
interaction generates a mass $m$ with
\eq (m\c)^2 = \alpha . \label{al} \eqend
We moved this mass term to the free part of the Hamiltonian so that
the Taylor series of the interaction starts with the forth order term,
\eq \cH^B_{1} = \sum_{n=2}^\infty \frac{(-1)^{n-1}(m\c)^2\beta^{2n-2}
 }{(2n)!} \int\dd x\, :\phi^{2n}: \; . 
\label{55} \eqend
In the non-relativistic limit we get, in leading order $1/m\c$,
\eq \phi = \frac{1}{\sqrt{2m} } \left( \Phi + \Phi^\dag + \ldots
\right)   \label{phi} \eqend
where $\Phi^{(\dag)}$ are non-relativistic boson fields obeying the
CCR $[\Phi(x),\Phi^\dag(y)]=\delta(x-y)$ (see Appendix~B.2). Thus the
coefficient in front of the $n$-th order term in the interaction is
$\propto m^{2-n} \beta^{2n-2} \c^2$, suggesting that, if the mass is
large, one only needs to take into account the leading term $n=2$ of
the interaction. We thus conclude that, for large values of $\alpha$,
the qSG model has the same non-relativistic limit as
$\phi_{1+1}^4$-theory. Using that we find that the qSG Hamiltonian, in
leading orders of $1/m$ and $1/\c$, reduces to
\eq \cH^B_{\rm non-rel} = \int\dd x\,
\frac1{2m}\Phi^\dag(x)(-\partial_x^2)\Phi(x) - \frac{(\beta\c)^2}{16}
: \Phi^\dag(x)\Phi(x) \Phi^\dag(x)\Phi(x) : \eqend
where normal ordering is defined with respect to the vacuum
$|0\rangle$ obeying $\Phi|0\rangle=0$ (see Appendix~B.2 for more 
details). This Hamiltonian now is well-defined, and for $2m=1$ and
$(\beta\c/4)^2=-\cB$ it is identical with the second quantization of
the 1D boson Hamiltonian in equation \Ref{Hp} (see equations  (1.1)--(1.12) in
\cite{Korepin}).

In Paragraph~4 we found a duality between the fermion Hamiltonian
defined in equation \Ref{H} and the boson Hamiltonian in equation \Ref{Hp}
with the relation of coupling parameters given in equation \Ref{dual}. It
is interesting to compare this to Coleman's duality between the
massive Thirring model and the qSG model (see equation (1.9) in \cite{C}),
\eq \frac{4\pi}{\beta^2} = 1 + \frac{g}{\pi} . \label{Co} \eqend
Inserting the relations $\lambda=-g/\c^2$ and $\cB=-(\beta\c/4)^2$
which we obtained in the non-relativistic limits in Paragraph~3 and
above, we obtain the relation in equation \Ref{dual} up to a factor
$\pi^2/4$ (the 1 on the r.h.s. in equation \Ref{Co} disappears in the
limit $\c\to\infty$). We regard this agreement up to a numerical
factor of order one as strong evidence that the the duality found in
Paragraph~4 is indeed the non-relativistic limit of Coleman's duality
(note that an exact agreement cannot be expected since we ignore the
renormalization of parameters in the qSG and massive Thirring models
\cite{C}). Note, however, that this argument only applies to the {\em
attractive} case $\lambda<0$, whereas the duality in equation \Ref{dual}
is true also for $\lambda>0$.

It is important to note that Coleman's duality provides also an
identification of field operators in the qSG and the Thirring models
(see equations  (1.10) and (1.11) in \cite{C}), but we do not see how this
identification appears in our non-relativistic limits. We therefore
regard the arguments in this paragraph only as a heuristic explanation
of the duality in equation \Ref{dual}. It would be interesting to
substantiate it in greater depth.

\bigskip

\sct{6. Generalization of the model to distinguishable particles.}  We
now discuss the generalized model with the Hamiltonian in equation \Ref{H}
but for {\em distinguishable} particles. We follow Yang \cite{Y} and
make the following Bethe Ansatz for the eigenfunctions,
\eq \chi = \sum_{P\in S_N} A_P(Q) \exp{\bigl( \ii \mbox{$\sum_{j=1}^N
k_{Pj} x_{Qj} $}\bigl) } \quad \mbox{for $x_{Q1} < x_{Q2} < \ldots
x_{QN}$,} \label{Bethe1} \eqend
for all $Q\in S_N$, which implies $E=\sum_j k_j^2$. It is
straightforward to adapt Yang's computation to our boundary conditions
in equation \Ref{BC}.  Inserting the Bethe Ansatz we obtain
\begin{align}
\ii(k_{Pi} - k_{P(i+1)}) [A_{PT_i}(QT_i) - A_P(QT_i)] &=
\ii(k_{Pi} - k_{P(i+1)}) [ A_P(Q)- A_{PT_i}(Q)]\\ 
A_{P}(QT_i) +
A_{PT_i}(QT_i) - A_P(Q) - A_{PT_i}(Q) &= 2\lambda \ii(k_{Pi} -
k_{P(i+1)}) [ A_P(Q) - A_{PT_i}(Q)] 
\end{align}
where $T_i$ is the transposition interchanging $i$ and $i+1$. This
implies
\eq (1 - \ii\lambda [ k_{P(i+1)} - k_{Pi}]) A_P(Q) = A_{PT_i}(QT_i) - 
\ii\lambda [ k_{P(i+1)} - k_{Pi}] A_{PT_i}(Q) . \eqend
We now use that the permutation group $S_N$ acts on the coefficients
$A_P(Q)$ by the regular representation $R\to \hat R$ as follows,
\eq A_P(QR) = (\hat R A_P) (Q)\equiv \sum_{Q'\in S_N} \hat R_{Q,Q'}
A_P(Q') \label{Z0} \eqend
with $\hat R_{Q,Q'} = \delta_{Q',QR}$. We thus can insert
$A_{PT_i}(QT_i) = (\hat T_i A_{PT_i})(P)$ and write this latter
relation as follows,
\eq
\label{Z1}
A_P = Y_i(k_{P(i+1)}-k_{Pi} ) A_{PT_i} \eqend
where $A_P$ here is a vector with $N!$ elements $A_P(Q)$ and 
\eq
\label{Z2}
Y_i(u) = \frac{\ii u \hat I - (1/\lambda)\hat T_i}{\ii u - 1/\lambda }
. \eqend
As explained by Yang \cite{Y}, consistency of the Bethe Ansatz is
equivalent to $Y_i(u)$ satisfying certain consistency conditions. The
first one (unitarity) is fulfilled in our case,
\eq Y_i(-u) Y_i(u) = \hat I ,\eqend
but the second one (Yang-Baxter relations) is not:
\eq Y_i(v) Y_{i+1}(u+v) Y_i(u) \neq Y_{i+1}(u) Y_i(v+u)
Y_{i+1}(v) . 
\label{Yang1}  \eqend
We conclude that the generalized model with distinguishable particles
is not exactly solvable by the coordinate Bethe Ansatz.

\bigskip

\sct{7. Final comments.} It is well-known that, in addition to the
delta-function interaction which has been studied extensively in the
context of integrable many-body systems, there are other local
interactions which are physically very different \cite{AGHH}. Recently
it was found that one particular such interaction leads to an exactly
solvable many-body system of fermions in 1D which has a remarkable
duality to the 1D boson gas \cite{CS}. In this paper we found a
natural physical interpretation of this fermion model: we showed that
the boundary conditions used to define the model in \cite{CS}
naturally arise from the $N$-body Hamiltonian in equation \Ref{H}
which describes particles with local, momentum dependent two-body
interactions. We also showed that this Hamiltonian arises as
non-relativistic limit of the massive Thirring model, and we argued
that the above-mentioned duality to the 1D boson gas comes from the
well-known duality of the Thirring model to the quantum sine-Gordon
model. We then proposed a generalization of this model where the
particles are distinguishable, but we found that, different from the
delta interactions case \cite{Y}, this model cannot be solved exactly
by the coordinate Bethe Ansatz.

As discussed in Chapter I.4 of \cite{AGHH}, quantum mechanical point
interactions in 1D leading to the boundary conditions in equation \Ref{bc}
have been studied extensively in the literature from a different point
of view, and apparently it has been interpreted as a
$\delta'$-interaction (see \cite{Seba}). The interpretation we give in
this paper is very different and, as we hope to have convinced the
reader, more natural.

We believe that our results show that, from a physical and
mathematical point of view, the model defined in equation \Ref{H} is
equally interesting as the delta-function interaction model given in
equation \Ref{Hp}. It thus would be worthwhile to explore this model
further, e.g., extend our results to the finite interval with suitable
boundary conditions etc.

\bigskip

\noindent
{\bf Acknowledgments.}  E.L. would like to thank Jens Hoppe and Antti
Niemi for asking stimulating questions, Jouko Mickelsson for helpful
comments, and Martin Halln\"as for checking some of our
formulas. E.L. would also like to thank the Erwin Schr\"odinger
Institute in Vienna for hospitality where this work was
started. C.P. was supported by {\em Deutsche Forschungsgemeinschaft}
under the Emmy-Noether programme. E.L. was supported in part by the
Swedish Science Research Council~(VR) and the G\"oran Gustafsson
Foundation.

\app \sct{Appendix~A: Physical interpretation of the interaction.}  In
this Appendix we give a complimentary physical interpretation of the
method to make sense of our momentum dependent interaction described
in Paragraph~2. For simplicity we restrict ourselves to the 2-particle
Hamiltonian $h$ in equation \Ref{H2}.

In the main text we gave a formal argument converting the interaction
in the Hamiltonian $h$ to the boundary conditions in equation \Ref{bc}.
It is interesting to note that, in doing this, we have specified a
regularization procedure, i.e., given a consistent prescription
avoiding divergences which would occur in a naive treatment of the
singular interaction. Indeed, naively the action of $h$ on a wave
function $\chi(x)$ is $(h\chi)(x) = -\chi''(x) +
4\lambda\delta'(x)\chi'(0)$, but from our discussion in Paragraph~2 it
is clear that $h$ is also defined on wave function which are
discontinuous at $x=0$ and with $\chi'(0)$ therefore undefined. The
above-mentioned regularization procedure amounts to replacing the
ill-defined derivate at $x=0$ by the well-defined average of the left-
and right derivatives at $x=0$, $\chi'(0)\to
[\chi'(0^+)+\chi'(-0^+)]/2$. To see that this eliminates a divergence
it is instructive to re-derive the bound state energy using Fourier
transformation. The Fourier transform of $h\chi=E\chi$ can be written
as
\eq (k^2-E)\hat\chi(k) = \lim_{\epsilon \to 0} 
4\lambda k \int_{\R} \frac{\dd q}{2\pi}
\cos(\epsilon q) q \hat \chi(q) \eqend
where the r.h.s.\ comes from the interaction with the factor
$\cos(\epsilon q)$ providing the regularization and the hat indicating
Fourier transform. Computing from this $\hat\chi(k)$, multiplying with
$k\cos(\epsilon k)$ and integrating we get the following
self-consistency relation,
\eq 1 = 4\lambda \lim_{\epsilon\to 0} 
\int_{\R} \frac{\dd q}{2\pi} \frac{\cos(\epsilon q)
q^2}{q^2 + |E|}   \eqend
where we used that the bound state energy is negative,
$E=-|E|$. Obviously, without the factor $\cos(\epsilon q)$ the
integral on the r.h.s.\ is linearly divergent, but with this factor we
obtain the well-defined result $1=-2\lambda\sqrt{|E|}$, which for
$\lambda=-|\lambda|$ has one solution. It is easy to see that this
yields the same value for the bound state energy and the same bound
state wave function which we obtained by a different method in
Paragraph~2.  \appende

\bigskip

\app \sct{Appendix B. Non-relativistic limits: Details.} In this
Appendix we give more details about how to derive the non-relativistic
limits of the Thirring model (Appendix~B.1) and $\phi^4_{1+1}$-theory
(Appendix~B.2) discussed in the main text.

\medskip
\noindent {\em B.1 Thirring model.} The Dirac Hamiltonian in Eq.\
\Ref{cH0} in Fourier space is
\eq \cH_0 = \int \dd k\; : (\hat \psi_+^\dag,\hat \psi_-^\dag )\left( \bma{cc}
k\c  - E_0 & m\c^2 \\ m\c^2 & -k\c - E_0 \ema
\right) \left( \bma{c} \hat \psi^\nd_+ \\ \hat \psi^\nd_- \ema \right):
\label{cH01} \eqend
with $\hat\psi^{(\dag)}_\pm \equiv \hat\psi^{(\dag)}_\pm(k)$ and the
hat indicating Fourier transform. It is diagonalized with the
following canonical transformation,
\eqa \hat \psi_\pm(k) = a_\pm(k) \hat \Psi_+(k) \pm a_\mp(k) \hat
\Psi_-(k)
\label{psipm}
\eqaend
where
\eq a_\pm(k) = \sqrt{\frac12\bigl(1\pm \frac{k\c}{E_k}\bigr) } , \quad
E_k = \sqrt{(m\c^2)^2 + (k\c)^2} .  \label{Ek} \eqend
This yields
\eq \cH_0 = \int\dd k\, : \Bigl( [E_k-E_0]\hat\Psi_+^\dag(k) \hat
\Psi_+(k) - [E_k+E_0]\hat\Psi_-^\dag(k) \hat \Psi_-(k) \Bigr): .
\eqend
Expanding this in powers of $k/m\c$ and transforming back to position
space one obtains the equations given in \Ref{Psipm} {\em ff} in the
main text. Transforming the interaction in equation \Ref{cHint} to Fourier
space, inserting the equations in \Ref{psipm}, and ignoring the terms
involving the negative energy fields $\hat\Psi^{(\dag)}_-$ we obtain
\eq \cH^+_{\rm int} = \frac{2g}{\pi} \int\dd k_1\cdots \dd k_4\,
\delta(k_1-k_2+k_3-k_4) v(k_1,\ldots,k_4) : \Psi^\dag(k_1) \Psi(k_2)
\Psi^\dag(k_3) \Psi(k_4): \label{cHint1} \eqend
with the interaction vertex
\eqa v(k_1,\ldots,k_4) = \frac{1}{4} \bigl( a_+(k_1) a_+(k_2) a_-(k_3)
a_-(k_4) + a_+(k_3) a_+(k_4) a_-(k_1) a_-(k_2)  \nonu - a_+(k_3)
a_+(k_2) a_-(k_1) a_-(k_4) - a_+(k_1) a_+(k_4) a_-(k_3) a_-(k_2)
\bigr) \eqaend
which we (anti-) symmetrized using the CAR. Expanding this in powers
of $1/m\c$ we obtain
\eq v(k_1,\ldots,k_4) = \frac{1}{(4m\c)^2}
 (k_1-k_3)(k_2-k_4) + O((m\c)^{-3}) . \eqend
Inserting this into equation \Ref{cHint1} and transforming back to
position space we obtain the interaction term in the non-relativistic
Hamiltonian given in equation \Ref{cHnonrel}.

\medskip

\noindent {\em B.2 $\phi^4_{1+1}$-theory.} This model can be formally
defined by the Hamiltonian $\cH^B=\cH^B_0 + \cH^B_{\rm int}$ with the
free part $\cH_0^B$ given in equation \Ref{cH0p} in the main text and the
interaction
\eq \cH^B_{\rm int} = g_B \int\dd x\, : \phi^4 :\label{33}\eqend
with boson fields $\phi$ and $\Pi$ as defined after equation \Ref{cH0p} in
the main text, $m>0$ the mass, and $g_B$ the coupling; the dots
indicate normal ordering to be specified below.  The free boson
Hamiltonian in equation \Ref{cH0p} can be diagonalized in the usual
manner,
\eqa \phi(x) = \c \int \frac{\dd k}{\sqrt{2\pi}}\frac1{\sqrt{2E_k}}
\left(\hat \Phi(k)\ee{\ii kx} + \hat \Phi^\dag(k)\ee{-\ii kx}
\right)\nonu \Pi(x) = -\frac{\ii}{\c}\int \frac{\dd
k}{\sqrt{2\pi}}\sqrt{\frac{E_k}{2}} \left(\hat \Phi(k)\ee{\ii kx} -
\hat \Phi^\dag(k)\ee{-\ii kx} \right) \label{22} \eqaend
with $E_k$ as in equation \Ref{Ek} and the $\hat \Phi^{(\dag)}$ the
Fourier transform of non-relativistic boson fields $\Phi^{(\dag)}$
obeying the CCR $[\Phi(x), \Phi^\dag(y)]=\delta(x-y)$ etc.  This
yields $\cH_0^B = \int\dd k \, E_k \hat \Phi^\dag (k) \hat \Phi(k)$
where, at this point, normal ordering is defined with respect to the
non-interacting vacuum $|0\rangle$ obeying
$\Phi(x)|0\rangle=0$. Expanding in powers of $1/m\c$ and transforming
to position space we get
\eq \cH^B_0 = \int\dd x\, \Phi^\dag(x)[m\c^2 - \partial_x^2/2m+
O((m\c)^{-1} )]\Phi(x) . \label{44} \eqend
To lowest non-trivial order in $1/m\c$ the first equation in \Ref{22}
reduces to equation \Ref{phi}.  Inserting this into the interaction in
equation \Ref{33} we get five terms, but only one of them commutes with
the particle number operator $\hat N = \int\dd x\,
\Phi^\dag(x)\Phi(x)$, namely $6 g_B/(2m)^2 \int\dd x\,
:[\Phi^\dag(x)\Phi(x)]^2:$. The other terms describe processes where
the particle number is changed, and since the creation of particles
requires an energy larger than $m\c^2$ (according to equation \Ref{44})
all these processes can be ignored in the non-relativistic limit where
$m\c$ becomes large.\footnote{While this is physically plausible, we
do not know a convincing mathematical argument to justify this
simplification. We therefore regard this step as the weak link in our
chain of arguments relating the SG model to the 1D boson gas.} Thus
the non-relativistic limit of $\phi^4_{1+1}$-theory can be described
the Hamiltonian
\eq \cH^B_{\rm non-rel} = \int\dd x\, 
\frac1{2m}\Phi^\dag(x)(-\partial_x^2)\Phi(x) + \frac{3g_B}{2 m^2}
: [\Phi^\dag(x)\Phi(x)]^2 : \eqend
which, for $2m=1$ and $3g_B/2 m^2 = \cB$, is the second quantization of
the 1D boson gas Hamiltonian given in equation \Ref{Hp}.

\appende


\bigskip


\begin{thebibliography}{99}
\bibitem{LL} E. H. Lieb and W. Liniger, Exact analysis of an
interacting Bose gas. I. The general solution and the ground state,
Phys. Rev. {\bf 130}, 1605 (1963)
%%CITATION = PHRVA,130,1605;%%
\bibitem{Korepin} V. E. Korepin, N. M. Bogoliubov\ and\ A. G. Izergin,
{\it Quantum inverse scattering method and correlation functions},
Cambridge Univ. Press, Cambridge (1993)
\bibitem{Y} C. N. Yang, Some exact results for the many-body problem
in one dimension with repulsive delta-function interaction,
Phys. Rev. Lett. {\bf 19}, 1312 (1967) 
%%CITATION = PRLTA,19,1312;%%
\bibitem{O} M. Olshanii, Atomic scattering in the presence of an
external confinement and a gas of impenetrable bosons,
Phys. Rev. Lett. {\bf 81}, 938 (1998)
\bibitem{G} M. Gaudin, Un systeme a une dimension de fermions en
interaction, Phys. Letters {\bf 24A}, 55 (1967)
\bibitem{FL} M. Flicker and E. H. Lieb, Delta-function fermi gas with
two-spin deviates, Phys. Rev. {\bf 161}, 179 (1967)
\bibitem{Th} W.~E.~Thirring, A soluble relativistic field theory,
Annals Phys.\ NY {\bf 3}, 91 (1958)
%%CITATION = APNYA,3,91;%%
\bibitem{CS} T. Cheon and T. Shigehara, Fermion-boson duality of
one-dimensional quantum particles with generalized contact
interactions, Phys. Rev. Lett. {\bf 82}, 2536 (1999)
%%CITATION = QUANT-PH 9806041;%%
\bibitem{C} S.\ Coleman, Quantum sine-Gordon equation as the massive
Thirring model, Phys. Rev. D {\bf 11}, 2088 (1975)
%%CITATION = PHRVA,D11,2088;%%
\bibitem{BT} H. Bergknoff and H. B. Thacker, Structure and solution of
the massive Thirring model, Phys. Rev. D {\bf 19}, 3666 (1979)
%%CITATION = PHRVA,D19,3666;%%
\bibitem{Smirnov} F.A. Smirnov, {\it Form Factors in Completely
Integrable Models of Quantum Field Theory}, Advanced Series in
Mathematical Physics, V. 14, World Scientific, Singapore (1992)
\bibitem{FW} L. L. Foldy and S. A. Wouthuysen, On the Dirac theory of
spin 1/2 particles and its non-relativistic limit, Phys. Rev. {\bf
78}, 29 (1950)
%%CITATION = PHRVA,78,29;%%
\bibitem{GGT} F. Gesztesy, B. Thaller, and H. Grosse, Efficient method
for calculating relativistic corrections for spin-1/2 particles,
Phys. Rev. Lett. {\bf 50}, 625 (1983)
%%CITATION = PRLTA,50,625;%%
\bibitem{Gaudin} M. Gaudin, {\it La fonction d'onde de Bethe}, Paris
  Masson (1983)
\bibitem{grouptheory} M. Hamermesh, {\it Group theory and its
application to physical problems}, Addison-Wesley Publishing Co.,
Inc., Reading, Mass. (1962)
\bibitem{YY} C. N. Yang and C. P. Yang, Thermodynamics of
one-dimensional system of bosons with repulsive delta function
interaction, J.\ Math.\ Phys.\ {\bf 10}, 1115 (1969)
%%CITATION = JMAPA,10,1115;%%
\bibitem{AGHH} S. Albeverio, F. Gesztesy, R. H{\o}egh-Krohn, and
H. Holden, {\it Solvable Models in Quantum Mechanics},
Springer-Verlag, Heidelberg (1988)
\bibitem{Seba} P. Seba, Some remarks on the $\delta'$-interaction in
one dimension, Rep. Math. Phys. {\bf 24}, 111 (1986)
\end{thebibliography}
\end{document}